\documentclass{aa}
\usepackage{psfig}
\begin{document}
%
%
\def\etal {et al.}
\def\ie {i.\,e.}
\def\etseq {{\em et seq.}}
\def\vs {{it vs.}}
\def\perse {{it per se}}
\def\adhoc {{\em ad hoc}}
\def\eg {e.\,g.}
\def\etc {etc.}
\def\ccpers {\hbox{${\rm cm}^3{\rm s}^{-1}$}}
\def\vlsr {\hbox{${v_{\rm LSR}}$}}
\def\vhel {\hbox{${v_{\rm HEL}}$}}
\def\delv {\hbox{$\Delta v_{1/2}$}}
\def\TL {$T_{\rm L}$}
\def\TC {$T_{\rm c}$}
\def\TEX {$T_{\rm ex}$}
\def\TMB {$T_{\rm mb}$}
\def\TKIN {$T_{\rm kin}$}
\def\TREC {$T_{\rm rec}$}
\def\TSYS {$T_{\rm sys}$}
\def\TVIB {$T_{\rm vib}$}
\def\TROT {$T_{\rm rot}$}
\def\TDUST {$T_{\rm d}$}
\def\TASTAR {$T_{\rm A}^{*}$}
\def\TVIBST {$T_{\rm vib}^*$} 
\def\H0 {$H_{\rm o}$}
\def\mic {$\mu\hbox{m}$}
\def\micro {\mu\hbox{m}}
\def\SDOZ {\hbox{$S_{12\mu \rm m}$}}
\def\STWE {\hbox{$S_{25\mu \rm m}$}}
\def\SSIX {\hbox{$S_{60\mu \rm m}$}}
\def\SHUN {\hbox{$S_{100\mu \rm m}$}}
\def\solmass {\hbox{M$_{\odot}$}}
\def\solum {\hbox{L$_{\odot}$}}
\def\irlum {\hbox{$L_{\rm FIR}$}}
\def\ohlum {\hbox{$L_{\rm OH}$}}
\def\blum {\hbox{$L_{\rm B}$}}
\def\numd {\hbox{$n({\rm H}_2$)}}                   
\def\rhounit {$\hbox{M}_\odot\,\hbox{pc}^{-3}$}
\def\kms {\hbox{${\rm km\,s}^{-1}$}}
\def\kmsyr {\hbox{${\rm km\,s}^{-1}\,{\rm yr}^{-1}$}}
\def\kmsmpc {\hbox{${\rm km\,s}^{-1}\,{\rm Mpc}^{-1}$}} 
\def\Kkms {\hbox{${\rm K\,km\,s}^{-1}$}}
\def\percc {$\hbox{{\rm cm}}^{-3}$}    
\def\cmsq  {$\hbox{{\rm cm}}^{-2}$}    
\def\cmsix  {$\hbox{{\rm cm}}^{-6}$}  
\def\arcsec {\hbox{$^{\prime\prime}$}}
\def\arcmin {\hbox{$^{\prime}$}}
\def\ffam {\hbox{$\,.\!\!^{\prime}$}}
\def\ffas {\hbox{$\,.\!\!^{\prime\prime}$}}
\def\ffM {\hbox{$\,.\!\!\!^{\rm M}$}}
\def\ffm {\hbox{$\,.\!\!\!^{\rm m}$}}
\def\ffs {\hbox{$.\,\!\!^{\rm s}$}}
\def\HI  {\hbox{\ion{H}{i}}}
\def\HII {\hbox{HII}}
%
%
\def \AL {$\alpha $}    
\def \BE {$\beta $}     
\def \GA {$\gamma $}    
\def \DE {$\delta $}    
\def \EP {$\epsilon $}  
\def \alde {($\Delta \alpha ,\Delta \delta $)}
\def \MU {$\mu $}       
\def \TAU {$\tau $}     
\def \tapp {$\tau _{\rm app}$}
\def \tuns {$\tau _{\rm uns}$}
\def \RH {\hbox{$R_{\rm H}$}}         
\def \RT {\hbox{$R_{\rm \tau}$}}      
\def \BN  {\hbox{$b_{\rm n}$}}        
\def \BETAN {\hbox{$\beta _n$}}       
\def \TE {\hbox{$T_{\rm e}$}}         
\def \NE {\hbox{$N_{\rm e}$}}         
%
\def\MOLH {\hbox{${\rm H}_2$}}                    
\def\HDO {\hbox{${\rm HDO}$}}                     
\def\AMM {\hbox{${\rm NH}_{3}$}}                  
\def\NHTWD {\hbox{${\rm NH}_2{\rm D}$}}           
\def\CTWH {\hbox{${\rm C_{2}H}$}}                 
\def\TCO {\hbox{${\rm ^{12}CO}$}}                 
\def\CEIO {\hbox{${\rm C}^{18}{\rm O}$}}          
\def\CSEO {\hbox{${\rm C}^{17}{\rm O}$}}          
\def\CTHFOS {\hbox{${\rm C}^{34}{\rm S}$}}        
\def\THCO {\hbox{$^{13}{\rm CO}$}}                
\def\WAT {\hbox{${\rm H}_2{\rm O}$}}              
\def\WATEI {\hbox{${\rm H}_2^{18}{\rm O}$}}       
\def\CYAC {\hbox{${\rm HC}_3{\rm N}$}}            
\def\CYACFI {\hbox{${\rm HC}_5{\rm N}$}}          
\def\CYACSE {\hbox{${\rm HC}_7{\rm N}$}}          
\def\CYACNI {\hbox{${\rm HC}_9{\rm N}$}}          
\def\METH {\hbox{${\rm CH}_3{\rm OH}$}}           
\def\MECN {\hbox{${\rm CH}_3{\rm CN}$}}           
\def\CH3C2H {\hbox{${\rm CH}_3{\rm C}_2{\rm H}$}} 
\def\FORM {\hbox{${\rm H}_2{\rm CO}$}}            
\def\MEFORM {\hbox{${\rm HCOOCH}_3$}}             
\def\THFO {\hbox{${\rm H}_2{\rm CS}$}}            
\def\ETHAL {\hbox{${\rm C}_2{\rm H}_5{\rm OH}$}}  
\def\CHTHOD {\hbox{${\rm CH}_3{\rm OD}$}}         
\def\CHTDOH {\hbox{${\rm CH}_2{\rm DOH}$}}        
\def\CYCP {\hbox{${\rm C}_3{\rm H}_2$}}           
\def\CTHHD {\hbox{${\rm C}_3{\rm HD}$}}           
\def\HTCN {\hbox{${\rm H^{13}CN}$}}               
\def\HNTC {\hbox{${\rm HN^{13}C}$}}               
\def\HCOP {\hbox{${\rm HCO}^+$}}                  
\def\HTCOP {\hbox{${\rm H^{13}CO}^{+}$}}          
\def\NNHP {\hbox{${\rm N}_2{\rm H}^+$}}           
\def\CHTHP {\hbox{${\rm CH}_3^+$}}                
\def\CHP {\hbox{${\rm CH}^{+}$}}                  
\def\ETHCN {\hbox{${\rm C}_2{\rm H}_5{\rm CN}$}}  
\def\DCOP {\hbox{${\rm DCO}^+$}}                  
\def\HTHP {\hbox{${\rm H}_{3}^{+}$}}              
\def\HTWDP {\hbox{${\rm H}_{2}{\rm D}^{+}$}}      
\def\CHTWDP {\hbox{${\rm CH}_{2}{\rm D}^{+}$}}    
\def\CNCHPL {\hbox{${\rm CNCH}^{+}$}}             
\def\CNCNPL {\hbox{${\rm CNCN}^{+}$}}             
%
%
\def\In {\hbox{$I^{n}(x_{\rm k},y_{\rm k},u_{\rm l}$})}
\def\Iobs {\hbox{$I_{\rm obs}(x_{\rm k},y_{\rm k},u_{\rm l})$}}
\def\Ingl {I^{n}(x_{\rm k},y_{\rm k},u_{\rm l})}
\def\Iobsgl {I_{\rm obs}(x_{\rm k},y_{\rm k},u_{\rm l})}
\def\Pbgl {P_{\rm b}(x_{\rm k},y_{\rm k}|\zeta _{\rm i},\eta _{\rm j})}
\def\Pbgm {P(x_{\rm k},y_{\rm k}|r_{\rm i},u_{\rm l})}
\def\Pbgn {P(x,y|r,u)}
\def\Pugm {P_{\rm u}(u_{\rm l}|w_{\rm ij})}
\def\Pdem {P_{\rm b}(x,y|\zeta (r,\theta ),\eta (r,\theta ))} 
\def\Pden {P_{\rm u}(u,w(r,\theta ))}
\def\greekgl {(\zeta _{\rm i},\eta _{\rm j},u_{\rm l})}
\def\greekg1 {(\zeta _{\rm i},\eta _{\rm j})}

\def\ffam {\hbox{$\,.\!\!^{\prime}$}}
\def\ffas {\hbox{$\,.\!\!^{\prime\prime}$}}
\def\ffM  {\hbox{$\,.\!\!^{\rm M}$}}
\def\ffm  {\hbox{$\,.\!\!^{\rm m}$}}

\def\JB   {\,Jy\,beam$^{-1}$}                 
\def\KB   {\,K\,beam$^{-1}$}                  
\def\MJB  {\,mJy\,beam$^{-1}$}                
\def\JBKS {\,Jy\,km\,s$^{-1}$\,beam$^{-1}$}   
\newcommand{\Rvmax}{$R_{\rm {\mbox{v$_{\rm max}$}}}$}
\newcommand{\rvmax}{R_{\rm {\mbox v_{\rm max}}}}
\newcommand{\Rmax}{$R_{\rm max}$}
\newcommand{\Vmax}{$V_{\rm max}$}
\newcommand{\vmax}{V_{\rm max}}
\newcommand{\Vsys}{$V_{\rm sys}$}        
\newcommand{\Vhel}{$V_{\rm hel}$}
\newcommand{\Vlsr}{$V_{\rm lsr}$}
\newcommand{\vrot}{$v_{\rm rot}$}        
\newcommand{\x}{\,$\times$\,}            
\newcommand{\HTWO}{H$_{\rm 2}$}          
\newcommand{\HA}{H$_{\rm \alpha}$}       
\newcommand{\MHTWO}{$M_{\rm H_{\rm 2}}$} 
\newcommand{\Msol}{M$_{\odot}$}          
\newcommand{\Lsol}{L$_{\odot}$}          
\newcommand{\LIR}{$L_{\rm IR}$}          
\newcommand{\Tsys}{$T_{\rm sys}$}        
\newcommand{\TAS}{$T^*_{\rm A}$}         
\newcommand{\COONE}{$^{12}$CO(1$-$0)}    
\newcommand{\COTWO}{$^{12}$CO(2$-$1)}    
\newcommand{\COTRI}{$^{12}$CO(3$-$2)}    
\newcommand{\LONE}{(1$-$0)}              
\newcommand{\LTWO}{(2$-$1)}              
\newcommand{\LTRI}{(3$-$2)}              
%

\title{ Atomic and molecular gas in the starburst galaxy NGC\,4945
         \thanks{Based on observations with the Australia Telescope 
          Compact Array. ATCA is funded by the Commonwealth of Australia
          for operation as a National Research Facility managed by 
          CSIRO.}
         \fnmsep
         \thanks{Based on observations with the Swedish-ESO submillimeter 
          telescope, ESO/La Silla, Chile}
       }

\author{M. Ott            \inst{1}
   \and J.B. Whiteoak     \inst{2}
   \and C. Henkel         \inst{1}
   \and R. Wielebinski    \inst{1}
       }

\institute{Max-Planck-Institute f\"ur Radioastronomie,
              Auf dem H\"ugel 69, D-53121 Bonn, Germany
      \and    Australia Telescope National Facility, CSIRO,
              PO Box Epping,
              NSW 2121, Australia        
             }

\offprints{R. Wielebinski}

\date{Received ... ; accepted ... }

\titlerunning{\hbox{\ion{H}{i}} and CO in NGC\,4945}

\authorrunning{M. Ott, J.B. Whiteoak, C. Henkel, and R. Wielebinski }


\abstract{
Spatial and kinematical correlations between the \hbox{\ion{H}{i}} and 
\COTWO\ emission of the southern spiral galaxy NGC\,4945 are studied with 
a common angular resolution of $\sim$\,23$''$ (corresponding to 750\,pc 
at $D=6.7$\,Mpc) and a velocity resolution of $\sim$7\,\kms. The 1.4\,GHz 
continuum emission is also observed. The \hbox{\ion{H}{i}} kinematics yield 
a galaxy mass of $\sim1.4$\x$10^{11}$\,\Msol\ within radius $R$=380$''$, 
with molecular and neutral atomic gas each contributing $\sim$2\%, respectively. 
A central continuum source of size 7\ffas6\x3\ffas4 (250\x110\,pc) is enveloped 
by a molecular cloud of mass $1.5\times10^9$\,\Msol\ for $R$ $\leq$ 7\ffas5, 
and is rapidly rotating with $V_{\rm rot}$ $\sim$ 160\,\kms. \hbox{\ion{H}{i}} 
emission from the central region at velocities $|V-$\Vsys$|$ $>$ 200\,\kms\ may 
be related to optically detected gas that is believed to trace an outflow directed 
towards the halo. Nuclear \hbox{\ion{H}{i}} absorption at $V$ -- \Vsys\ $\sim$ 
+80\,\kms\ suggests inflow towards the centre, that was so far only seen in 
molecular lines. \hbox{\ion{H}{i}} features at each end of the major axis 
($|R|$ $\sim$ 600$''$) are interpreted as spiral arms that are viewed tangentially 
and that also cause prominent emission features in the radio continuum, 
\hbox{\ion{H}{i}}, and CO further inside the galaxy. A central elongated region 
showing non-circular motions is interpreted as a bar which fuels the nuclear starburst. 
The \hbox{\ion{H}{i}} and CO position-velocity data have been analysed using linear 
resonance theory, and possible locations of resonances are identified. 
         }

\maketitle


\section{Introduction}

Studies of the formation and evolution of spiral galaxies require knowledge of
their morphological and kinematical properties. Atomic and molecular hydrogen
(\hbox{\ion{H}{i}} and \HTWO) are the main interstellar gas components of spiral 
galaxies and provide excellent tracers to elucidate spiral structure and rotation.
Since it is difficult to directly observe \HTWO, carbon-monoxide (CO) is commonly 
used to probe \HTWO\ column densities and molecular masses. Whereas \hbox{\ion{H}{i}} 
is a good tracer for the outer regions of spiral galaxies, the inner regions are 
often better studied in CO.

The starburst galaxy NGC\,4945 is particularly suited for high resolution
and high sensitivity studies. Being a member of the Centaurus group, the nearby
edge-on galaxy contains a Seyfert 2 nucleus and is classified as SB(s)cd or 
SAB(s)cd (de Vaucouleurs 1964; Braatz et al. 1997). Distance estimates range 
from 3.8\,Mpc (de Vaucouleurs 1964; Bergman et al. 1992) to 8.1\,Mpc (Baan 
1985). In order to be consistent with other relevant studies, throughout 
the paper 6.7\,Mpc is used.

Although at optical wavelengths the active nucleus is obscured ($N_{\rm H}$ 
$\ga$ 10$^{24}$\,\cmsq; Guainazzi et al. 2000; Madejski et al. 2000), there 
is evidence for a nuclear superwind (Chen \& Huang 1997; Lipari et al. 
1997) and, at $\lambda$ = 100\,$\mu$m, the central region is one of the three 
brightest IRAS point sources beyond the Magellanic Clouds (IRAS 1989). The 
nucleus is well-defined at radio frequencies ($S_{\rm 1.4GHz}$ = 4.9\,Jy; 
Elmouttie et al. 1997) and shows little ultra compact structure on a 
milliarcsec scale (Preston et al. 1985; Sadler et al. 1995). Microwave 
transitions of molecules have been detected in dense clouds that envelope 
the nucleus; this includes the first discovered `megamaser' (Dos Santos 
\& L\'epine 1979; Batchelor et al. 1982; Whiteoak \& Gardner 1986), a 
possible circumnuclear disk of radius 0.3\,pc with a binding mass of 
$\sim$10$^{6}$\,\solmass\ (Greenhill et al. 1997), and numerous transitions 
at cm and mm wavelengths (see e.g. Whiteoak 1986; Henkel et al. 1990, 1994; 
Curran et al. 2001). CO studies suggest the presence of a nuclear gas ring 
(Whiteoak et al. 1990; Bergman et al. 1992; Dahlem et al. 1993). Properties 
of NGC\,4945 relevant to this paper are summarized in Table 
\ref{TAB.INT.PROPERTIES}.

\begin{table}[t]
\caption[Known properties of NGC\,4945]{Properties of NGC\,4945
(9 mag, SB(s)cd)\\}
{\footnotesize
\begin{center}
\label{TAB.INT.PROPERTIES}
\begin{tabular}{llll}
\hline	
Position              & $\alpha${\tiny (2000)} & 13$^{\rm h}$ 05$^{\rm m}$
                                                 27\ffs 4             &    \\
~~~~of nucleus        & $\delta${\tiny (2000)} & --49$\degr$ 28$'$ 05$''$
                                                                    & 1, 2 \\
Distance$^{\rm a)}$   &  $ D $         & 6.7\,Mpc                   & 3    \\  
Diameter              &  $D_{\rm 25m}$ & $17' $                     & 4    \\
Systemic              & \Vhel          & 563 and 561$\pm$4\,\kms    & 5, 7 \\
~~~~velocity$^{\rm b)}$ & \Vlsr  & 555\,\kms                        & 6    \\
Inclination           &   $ i $        & $78 \pm 3\degr $           & 6, 7 \\
Position angle        &   $ PA $       & $43\pm1\degr$ ; $45\pm2\degr$ &6, 7\\
Turn-over radius      &   \Rvmax       & $5\farcm45$ ;  $6\farcm6$  & 6, 7 \\
Turn-over-velocity    & \Vmax          & 165 and  $183\pm4$\kms     & 6, 7 \\
Total mass            &  $M_{\rm T}$   & 8.8 and 15 $\times$ $10^{10}$\Msol 
                                                                    & 6, 7 \\
\hline
\end{tabular}
\end{center}
}
{\footnotesize
a) The value of 6.7\,Mpc was chosen for consistency with previous \hbox{\ion{H}{i}} 
and CO studies; at this distance 23$''$ corresponds to 750 pc. Evidence for $D=4$\,Mpc 
has been summarized by Bergman et al. (1992). If this value were used, gas 
masses and luminosities have to be multiplied by a factor of 0.36 except for 
dynamical masses, for which a factor of 0.6 is appropriate.\\
b) \Vhel\ $=$ \Vlsr\ + 4.7\,\kms\\
\begin{tabular}{ll}
References: &\\
1) Whiteoak \& Bunton (1985)    &  5) Whiteoak et al. (1990) \\
2) Elmouttie et al. (1997)      &  6) Ables et al. (1987)    \\ 
3) Whiteoak \& Gardner (1977)   &  7) Dahlem et al. (1993)   \\
4) de Vaucouleurs et al. (1991) \\
\end{tabular}
}
\end{table}

Early optical spectroscopic observations suggest that the rotational velocity 
increases linearly out to a radius of $r=360''$ but also show some evidence 
of non-circular motions (Peterson 1980; Carranza \& Ag\"uero 1983). Studies 
of the disk radio continuum, \hbox{\ion{H}{i}}, and CO emission have been made 
by Elmouttie et al. (1997), Ables et al. (1987), and Dahlem et al. (1993) 
respectively. The large-scale radio emission has a steep (frequency) spectral index 
($\alpha$ = --1.1$\pm$0.1); no emission was detected from the halo. \hbox{\ion{H}{i}} 
was found to extend out to $r=7'$ with an additional feature at $10'$ in the 
south-west. CO\LONE\ emission is concentrated towards the nucleus, but was 
also detected out to $6'$ along the major axis.

We have used the Compact Array of the  Australia Telescope National Facility
(ATCA, see Frater \& Brooks 1992) to study the 1.4\,GHz continuum and 
\hbox{\ion{H}{i}} emission of NGC\,4945 in more detail and with higher sensitivity 
than before. We have also observed CO\LTWO\ emission using the Swedish-ESO Submillimetre 
Telescope (SEST; see Booth et al. 1989). In addition to the results presented 
by Dahlem et al. (1993) our CO\LTWO\ data are not confined to the nuclear region 
but also include significant parts of the disk. \hbox{\ion{H}{i}} and CO have been 
compared at a common angular resolution of $\sim$23$''$ and channel spacings
of 6.6 and 6.3\,\kms, corresponding to velocity resolutions of $\sim$7.25 
and 6.5\,\kms, respectively; 23$''$ corresponds to 750\,pc at $D$ = 6.7\,Mpc 
(see also Table \ref{TAB.INT.PROPERTIES}, footnote `a').

\begin{figure*}
\psfig{figure=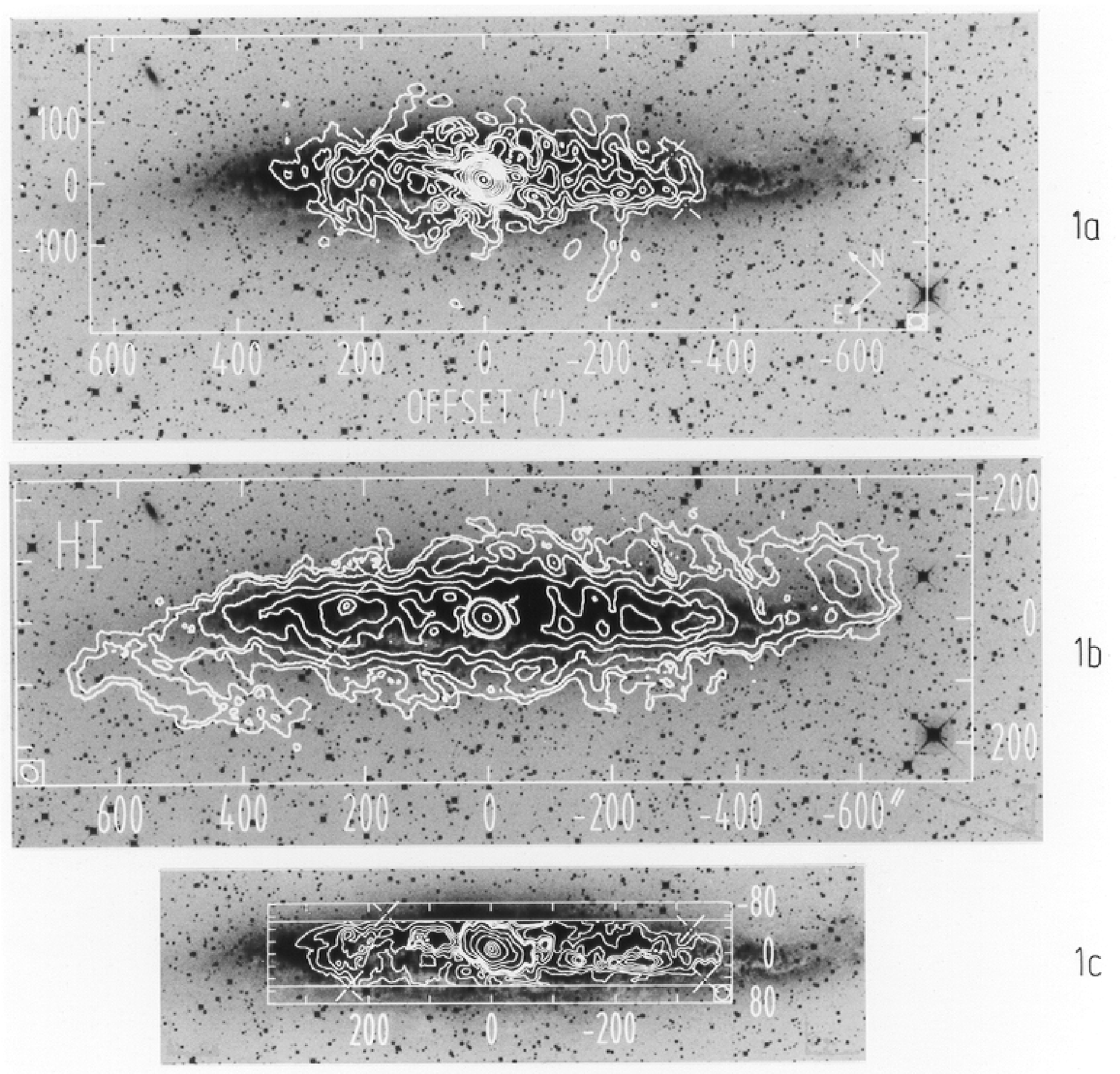,height=19cm}
\caption[Continuum on optical image]{ {\bf (a)} 1.4\,GHz continuum 
superimposed on a UK Schmidt Telescope SRC (Science Research Council; now
SERC = Science and Engineering Research Council) J (yellow sensitive 
emulsion/filter combination) image. The contour levels have flux densities 
of 5, 7.5, 10, 15, 20, 25, 30, 40, 60, 120, 250, 500, 1000, 2000, 3000, 
4200\MJB. The resolution is 19$''$ (R.A.) \x 25$''$ (Dec.) (i.e. an average 
of $\sim$23$''$) along major and minor axes. {\bf (b)} Integrated 
\hbox{\ion{H}{i}} emission superimposed on an SRC-J image. The contours 
denote $-$5000, $-$500, 3, 7.5, 15, 30, 50, 70, 90, 99\% of the integrated 
emission peak of 3.8\,\JBKS. The resolution is as in (a). {\bf (c)} 
Distribution of CO(2$-$1) emission integrated over velocity superimposed 
on a SRC-J image. The contours are 10, 20, 40, 80, 160, 320, 640, 
1280\,K\,\kms\,beam$^{\rm -1}$\ on a \TMB\ scale. The resolution is 
$\sim$23$''$, and the rms noise is $\sim$ 3\,K\,\kms\,beam$^{\rm -1}$. 
The four white crosses served as reference positions for the overlay.
}
\label{FIG.MAPS}
\end{figure*}

\section{Observations and data reduction}

\subsection{1.4\,GHz continuum emission and \hbox{\ion{H}{i}} synthesis data}

The ATCA was equipped with 20\,cm FET receivers  with system temperatures
between 33 and 55\,K. Four observing runs provided 53 independent baselines, 
and were sufficient to achieve a satisfactory sampling of the {\it u-v} plane. 
Table \ref{TAB.HI.CONFIG} displays the configurations, observing dates and 
baseline ranges. A standard correlator setup of 512 channels was used with 
a total bandwidth of 8\,MHz centred on 1418\,MHz. The channel separation was 
15.6\,kHz (3.3\,\kms). An observing cycle was adopted in which a 40-minute 
tracking of NGC\,4945 was bracketed  by alternate 5-min observations of the 
phase-calibration sources PKS\,1215-457 and PKS\,1320-446. The radio galaxy 
PKS\,1934-638 was observed as a flux-density calibrator (assumed to have a 
flux density of 16.4\,Jy at 1.4\,GHz). Observations of the quasar PKS\,0407-658 
provided antenna spectral bandpasses.

{\small
\begin{table}[h]
\caption[ATCA  configurations \& baselines]{
ATCA configurations for the \hbox{\ion{H}{i}} observations.}
\label{TAB.HI.CONFIG}
\begin{center}
\begin{tabular}{llcrrr}
\hline
\multicolumn{5}{l}{ } \\
Config.  &   Date                     &
\multicolumn{3}{l}{Baselines [m]}                \\
\hline
1.5\,B     &  Oct.\,18,\,1991  &     31 &   to &  4301 \\   
1.5\,C     &  Dec.\,3,\,1991   &     77 &   to &  4500 \\ 
0.75\,C    &  Aug.\,22,\,1992  &     46 &   to &  5020 \\
6.0\,C     &  Feb.\,12,\,1993  &    153 &   to &  6000 \\
\hline
\end{tabular}
\end{center}
\end{table}
}

The data were reduced with an ATNF modification (Killeen 1992) of the 
Astronomical Image Processing System (AIPS) of the US National Radio Astronomy
Observatory. A single data file of correlated {\it u-v} data was produced 
for each observing run. After editing of the spectra obtained with individual 
baselines for phase and amplitude errors, the central 400 spectral channels 
were selected.  Those channels free of \hbox{\ion{H}{i}} contributions were 
averaged in the {\it u-v} domain for each of the four data sets and merged into 
a continuum-emission data base. Standard imaging and CLEAN (H\"ogbom 1974) 
procedures were then applied, and the resultant image was further improved 
by a standard self-calibration routine (e.g. Cornwell \& Fomalont 1989).

In the {\it u-v} domain, the continuum emission was removed from channels
containing the \hbox{\ion{H}{i}} data using the line-free channels (van Langevelde 
\& Cotton 1990; Cornwell et al. 1992). After concatenation of the four 
\hbox{\ion{H}{i}} data sets, the gain solutions obtained from the self-calibration 
of the continuum image were applied.

To facilitate subsequent analysis two sets of images were produced for both 
the continuum and line data. `Natural weighting' provided a resolution of
19$''$ (R.A.) $\times$ 25$''$ (Dec.) for investigation of the extended 
structure; `uniform weighting' yielded a restoring beam of 3.2$''$ $\times$
4$''$ to reveal the spatial fine structure of the nuclear region. A correction 
for the gain variation across the ATCA 22-m antenna beams was applied to all 
images.

The rms noise of the final continuum image is $\sim2.0$\MJB. Faint artifacts 
are present with flux densities of 5 -- 8\MJB. The peak-to-noise ratio for 
the nucleus is of the order of 800. Averaging two contiguous channels (this
yields a channel spacing of 6.6\,\kms) the rms noise of the \hbox{\ion{H}{i}} 
data becomes 1.2\MJB.

To investigate the possibility of missing extended emission, the galaxy was 
also mapped with the Parkes 64-m telescope. The total \hbox{\ion{H}{i}} flux 
density integrated over velocity for Parkes observations (64\,Jy\,\kms ) was 
found to be 9\% lower than its ATCA counterpart (70\,Jy\,\kms ). The difference 
is within the uncertainty of at least 5\% for each estimate. In summary, 
the comparison of flux densities provides no evidence for missing emission 
in the ATCA observations.

\subsection{$^1$$^2$CO J=2$\rightarrow$1 data}

CO\LTWO\ line emission was mapped during 1993 March 1--10 with the SEST at an
angular resolution of 23$''$. A 230\,GHz SIS receiver yielded system 
temperatures of 600 -- 1300\,K on an antenna temperature (\TAS) scale. 
The main beam efficiency was 0.46. An acousto-optical spectrometer, with 
1440 channels and a total bandwidth of 1\,GHz, provided a channel separation 
of 0.9\,\kms.

A rectangular grid of positions was selected, centred on the nucleus of the
galaxy (see Table \ref{TAB.INT.PROPERTIES}), and with offsets parallel or
orthogonal to the major axis at $PA = 43\degr$. The spectra were obtained at
intervals of 10$''$ along the major and minor axes, and 14$''$ elsewhere. In
total, 404 positions were sampled with offsets from the nucleus that ranged
from $-370''$ to $+360''$ parallel to the major axis, and from $-60''$ to
$+60''$ perpendicular to it.

All spectra were obtained using a dual beam-switching mode (switching
frequency 6\,Hz), with a beam throw of 11\ffam 7. The integration time was
4\,min per position and, averaging seven contiguous channels (channel
spacing: 6.3\,\kms), the rms noise ranged between 20 and 45\,mK on a
\TAS\ scale. A set of three spectra was preceded by a short calibration 
observation of a black body `paddle' which provided an intensity conversion 
to \TAS. Periodic continuum observations of the small-diameter nucleus of 
the radio source Cen\,A at 115\,GHz provided antenna pointing corrections. 
The pointing was also assessed by regularly re-observing the profile shape 
of the CO spectrum at the central position of NGC\,4945. These measurements
infer absolute positional uncertainties $<$8$''$ and relative positional 
errors $<$5$''$. 

The data were processed with the CLASS package of the Groupe d'Astrophysique
de Grenoble. To facilitate a comparison of CO with \hbox{\ion{H}{i}}, the CO 
spectra were transformed into a data cube and transferred to the AIPS software.

\begin{figure}
\psfig{figure=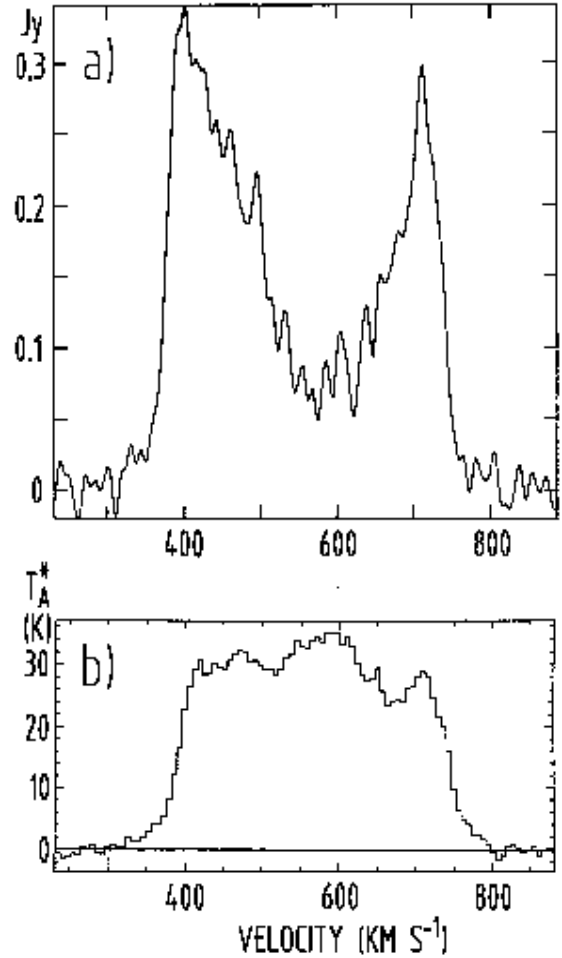,height=13cm}
\caption[\hbox{\ion{H}{i}} line profile]{ Line emission profiles as a function of 
Local Standard of Rest (LSR) velocity of {\bf (a)} the \hbox{\ion{H}{i}} and 
{\bf (b)} the CO\LTWO\ emission integrated over the observed region with channel 
spacings of 6.6 and 6.3\,\kms, respectively.
}
\label{FIG.HI.INTSPEC}
\label{FIG.CO21.TOTLINE}
\end{figure}

\section{Results and Discussion}

\subsection{A brief overview}

The distributions of 1.4\,GHz continuum emission, integrated \hbox{\ion{H}{i}} 
emission, and integrated CO\LTWO\ emission are shown in Fig.\,\ref{FIG.MAPS}, 
superimposed on an optical image from the UK Schmidt SRC (Science Research Council) 
survey. The position offsets are relative to $\alpha(2000)$ = 13$^{\rm h}$ 05$^{\rm m}$ 
27\ffs 4, $\delta(2000)$ = --49$\degr$ 28$'$ 05$''$.

\subsubsection{The radio continuum}

The continuum emission (Fig.\,\ref{FIG.MAPS}a) shows a bright small-diameter
nucleus superimposed on an elongated distribution of emission. The map is in
reasonable agreement with the 1.4\,GHz image of Elmouttie et al. (1997)
but shows additional details. The peak flux density of the central source is
4.2$\pm0.1$\JB\ at $\alpha(2000) = 13^{\rm h} 05^{\rm m}$ 27\ffs 3, 
$\delta(2000) = -49\degr 28' 07''$ (the position agrees to within $2''$ 
with those given by Whiteoak \& Bunton (1985) and Elmouttie et al. (1997)). 
The flux density integrated over the central source is 4.6$\pm$0.1\,Jy.

The emission extends over 11.6$'$ $\times$ 3.3$'$ with the major axis at a 
position angle $PA\sim45\degr$. Northeast of the nucleus a ridge-line runs 
above the major axis and veers to the east at $R$ $\sim$ 240$''$. Southwest 
of the nucleus the ridge appears to run below the major axis, turning to the 
west at $R$ $\sim$ 300$''$. Faint (5\,mJy) curved structures extending out 
of the disk at $R$ $\sim$ 150--200$''$ are residual sidelobes caused by the 
relatively bright emission at the nucleus (cf. Sect.\,2.1).

\begin{figure}
\psfig{figure=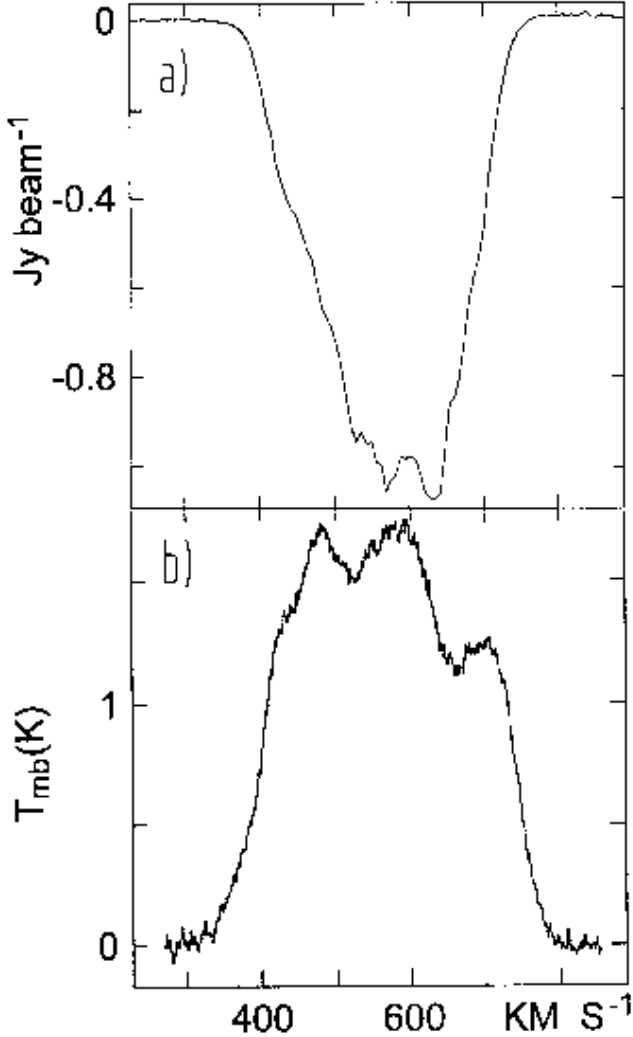,height=14cm}
\caption[Nuclear \hbox{\ion{H}{i}} absorption]{ {\bf (a)} \hbox{\ion{H}{i}} 
spectrum (LSR velocities) towards the central continuum source, obtained with 
an angular resolution of 19$''$ (R.A.) $\times$ 25$''$ (Dec.) and a channel 
spacing of 6.6\,\kms. {\bf (b)} The \COTWO\ line spectrum towards the centre 
position with an angular resolution of $23''$. The integration time was 60 
minutes and the channel spacing is 0.9\,\kms. 
}
\label{FIG.HI.CENABS}
\label{FIG.CO21.CENTERPOS}
\end{figure}

\subsubsection{The \hbox{\ion{H}{i}} distribution}

In our contour map of the \hbox{\ion{H}{i}} distribution (Fig.\,\ref{FIG.MAPS}b) 
the central region is dominated by \hbox{\ion{H}{i}} absorption against the strong 
continuum emission of the nucleus. However, no \hbox{\ion{H}{i}} absorption was 
observed against the extended continuum outside this central region. Because of 
the influence of the absorption, the total integrated flux density discussed 
in Sect.\,2.1 is not representative of the total \hbox{\ion{H}{i}} content. In 
the figure, the peak value of the integrated emission is 
$\sim3.8$\,Jy\,\kms\,beam$^{-1}$ (at $R$ $\sim$ +220$''$). The \hbox{\ion{H}{i}} 
emission extends over $\sim$ 22$'$ $\times$ 4$'$ (43 $\times$ 7.8\,kpc), showing 
a stretched `S'-structure. Within this, a ridge of emission extends $13'$ across 
the nucleus at $PA\sim 43^{\circ}$, whereas the major axis as defined by the 3\% 
intensity level is at $PA$ $\sim$ 50$^\circ$. At the south-western end of the 
distribution there is a moderately bright \hbox{\ion{H}{i}} concentration that 
is extending to the north. At the opposite end of the galaxy the \hbox{\ion{H}{i}} 
distribution extends to the south.

Fig.\,\ref{FIG.HI.INTSPEC}a shows the average \hbox{\ion{H}{i}} spectrum for the 
entire galaxy. The velocity relative to the Local Standard of Rest (LSR) extends 
from 340 to 770\,\kms\ and the emission peaks at 400 and 715\,\kms. The 
systemic velocity, as given by the average of the extreme and peak velocities, 
is \Vsys $\sim$ 555 and 555--560\,\kms, respectively (cf. Table 
\ref{TAB.INT.PROPERTIES}). The deep asymmetric central depression reflects 
the presence of absorption against the nucleus. The shape of the spectrum 
is similar to that obtained with the Parkes telescope directed towards the 
centre of the galaxy (Whiteoak \& Gardner 1977).

Fig.\,\ref{FIG.HI.CENABS}a shows our \hbox{\ion{H}{i}} spectrum with $\sim$23$''$ 
resolution towards the position of the nuclear continuum source. \hbox{\ion{H}{i}} 
is present in absorption against the continuum emission at velocities of 350 -- 
770\,\kms, i.e. symmetrically  offset up to $\sim$210\,\kms\ from \Vsys. The 
profile is asymmetric and appears to be composed of several components. A peak 
absorption flux density of 1.05\JB\ is observed at 570 and 635\,\kms. The 
corresponding line-to-continuum ratio is 0.25.

\begin{figure}
\psfig{figure=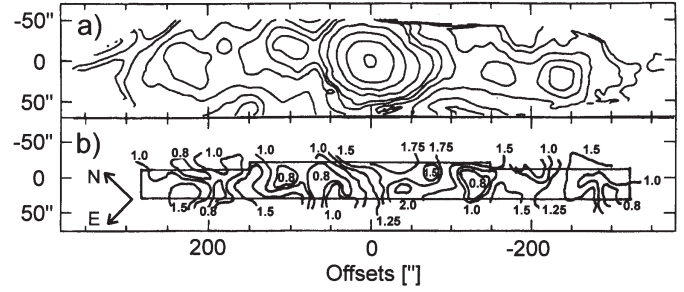,height=4cm}
\caption[CO(2$-$1) at 43$''$ and ratio map] {{\bf (a)} Integrated CO(2$-$1)
emission convolved to 43$''$. Contours are 5, 10, 20, 30, 40, 50, 80, 160,
320, 640\KB\ on a \TMB\ scale. {\bf (b)} The CO (2$-$1)/CO(1$-$0) ratio (in
units of \TMB) across the disk. The inner box denotes the approximate boundary
for `reasonably' convolved CO(2$-$1) spectra.
}
\label{FIG.CO21.AT43}
\end{figure}

\subsubsection{CO emission}

The distribution of CO\LTWO\ emission integrated over velocity 
(Fig.\,\ref{FIG.MAPS}c) extends over more than 10$'$ along the major axis. As 
the continuum and \hbox{\ion{H}{i}} distributions, CO also shows a ridge extending 
above the major axis north-east of the nucleus and below the major axis to 
the south-west. Like the continuum the CO distribution bends towards the 
east at $R$ $\sim$ 250$''$. 

Fig.\,\ref{FIG.CO21.TOTLINE}b shows the integrated CO(2$-$1) spectrum for the
entire region observed. The CO velocity range is about the same as for the 
\hbox{\ion{H}{i}} in Fig.\,\ref{FIG.CO21.TOTLINE}a. Broad peaks are present 
near 415, 480, 580, and 705\,\kms. For an assumed symmetric system, the outer 
peaks would suggest a systemic velocity of 560\,\kms.

Fig.\,\ref{FIG.CO21.CENTERPOS}b shows the CO\LTWO\ emission towards the 
nucleus of the galaxy. The CO covers the same velocity range as the 
previous spectrum. The profile shape is similar, except that the higher
velocity feature is fainter than the systemic feature. In our 
Fig.\,\ref{FIG.CO21.TOTLINE} and in Fig.\,5 of Dahlem et al. (1993),
both features have almost the same line temperature. This likely reflects 
small differences (a few arcsec) in the pointing of the telescope (cf. 
Sect.\,2.2). A fitting of gaussian components yields distinct components 
centred at velocities of 447, 493, 593, and 701\,\kms, with a further underlying
broad component centred at $\sim$565\,\kms. 

Fig.\,\ref{FIG.CO21.AT43}a shows the integrated CO(2$-$1) emission convolved
to a resolution of 43$''$. Combined with corresponding SEST CO(1$-$0) data
observed at the same resolution (Dahlem et al. 1993), the distribution of 
the CO(2$-$1)/CO(1$-$0) ratio is shown in Fig.\,\ref{FIG.CO21.AT43}b. The
ratio varies from 0.8 to 2.0 and demonstrates that `warm spots' with ratios
larger than unity are not confined to the central region but are also observed 
far out in the disk. For a possible spatial correlation of these warm molecular
regions with spiral arms, see Sect.\,3.4.

\begin{figure}
\psfig{figure=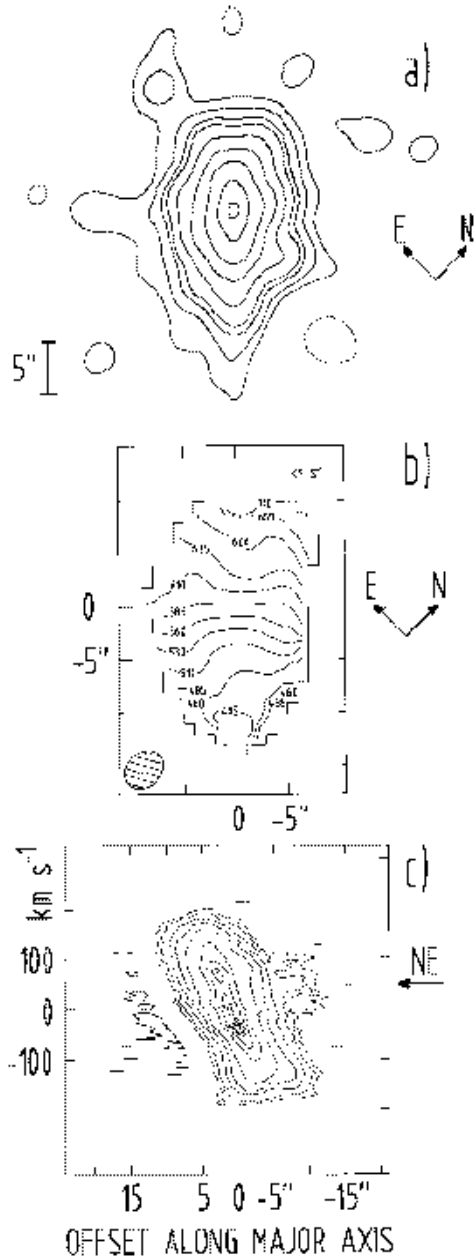,height=18cm}
\caption[Nuclear region at high resolution of 4$''$ resolution]{ The
resolved nucleus at a resolution of 3.2$''$ in R.A. and 4\ffas0 in Dec. 
{\bf (a)} The nuclear 1.4\,GHz continuum. The contour levels have flux 
densities of --2 (dashed), 10, 20, 40, 60, 120, 240, 500, 1000, 1300\MJB. 
{\bf (b)} The LSR velocity field of the nuclear \hbox{\ion{H}{i}} absorption 
with iso-velocity contours of 435, 460, 485, ...\ 685, and 710\,\kms. The 
vertical direction is equivalent to $PA=43^{\circ}$. {\bf (c)} Position-velocity 
diagram of the nuclear \hbox{\ion{H}{i}} absorption along $PA=43^{\circ}$ at 
$\sim$3.6$''$ $\times$ 7.25\,\kms\ resolution. The velocity axis is labeled 
with respect to 560\,\kms. Contours are -1, 1, 2,5, 10, 20, 50, 70, 85, 90, 
95, 99\% of the peak absorption of 612\,mJy (for the peak flux density in a 
$\sim$23$''$ beam, see Fig.\,\ref{FIG.HI.CENABS}a).
}
\label{FIG.NUC.HIRES}
\end{figure}

\subsection{The central region}

The strong \hbox{\ion{H}{i}} absorption against the central radio continuum 
source complicates a direct comparison of \hbox{\ion{H}{i}} and CO. The 
\hbox{\ion{H}{i}} absorption must originate from in front of the continuum 
whereas the CO emission may arise from in front of {\it and} behind the nucleus. 
This difference is consistent with the CO and \hbox{\ion{H}{i}} lineshapes 
shown in Fig.\,\ref{FIG.CO21.CENTERPOS}, where the CO profile is wider than 
its \hbox{\ion{H}{i}} counterpart at the half-maximum-intensity points. This 
effect must be significant: We could also plot, instead of the \hbox{\ion{H}{i}} 
absorbing flux, the \hbox{\ion{H}{i}} optical depth $\tau$(\hbox{\ion{H}{i}}) = 
--ln\,(1 + $S_{\rm L}$/$F\,S_{\rm c}$) with $F$ denoting the continuum source 
covering factor and $S_{\rm L}$ and $S_{\rm c}$ being the (negative) line and 
(positive) continuum flux. For $F$=1 the profile would resemble that shown in 
Fig.\,\ref{FIG.HI.CENABS}a, since the line remains optically thin even at the 
line centre. With $F$ $\sim$ 0.25, however, \hbox{\ion{H}{i}} optical depths 
would be large near $V$ $\sim$ 600\,\kms\ and the \hbox{\ion{H}{i}} column 
density profile would become narrower. The velocity of the central CO peak 
lies between the velocities of the two strongest \hbox{\ion{H}{i}} absorption 
components.
 
Fig.\,\ref{FIG.NUC.HIRES}a shows the nuclear continuum emission at a resolution 
of 3\ffas2 in R.A.\ and 4\ffas0 in Dec. It is spatially resolved: a gaussian fit 
and a correction for beam size yields a source size of 7\ffas6 $\times $ 3\ffas4 
$(\pm0\ffas 2)$ (corresponding to 250\x110\,pc), with the major axis at $PA=42\fdg5$ 
and a peak intensity of 1.33$\pm0.03$\MJB. The angular dimensions are reasonably 
consistent with those (5\ffas7 $\times $ 2\ffas0 $(\pm0\ffas1)$) determined by 
Whiteoak \& Wilson (1990) at 6\,GHz. Figs.\,\ref{FIG.NUC.HIRES}b and c show the 
velocity field of the \hbox{\ion{H}{i}} absorption towards the resolved nucleus 
and the associated position-velocity diagram. The velocity field is centred within 
15\,\kms\ of the galaxy systemic velocity (555--560\,\kms) and the velocity gradient 
across the nucleus is reasonably uniform. It supports the existence of a cloud which 
surrounds the nucleus and has rapid solid-body rotation. These results are consistent 
with the 6\,GHz OH study by Whiteoak \& Wilson (1990). However, because the 
\hbox{\ion{H}{i}} absorption is associated only with the region overlying the continuum 
emission, it cannot provide information about the overall size of the cloud. 

The strong CO\LTWO\ emission associated with the nucleus (cf. 
Fig.\,\ref{FIG.MAPS}c) has an elongated distribution. Its extent, derived
from a gaussian fit and corrected for beam size, is $37''$\x$21''$ with
$PA=13\degr$. Table\,\ref{TAB.CO.NUCSIZE} contains a comparison with
results for CO\LONE\ emission (Dahlem et al. 1993) and CO\LTRI\ emission 
(Mauersberger et al. 1996a). The emission associated with the highest-state 
$J$ = 3--2 transition is more centrally concentrated than for the 
lower-excitation transitions.

\begin{figure}
\psfig{figure=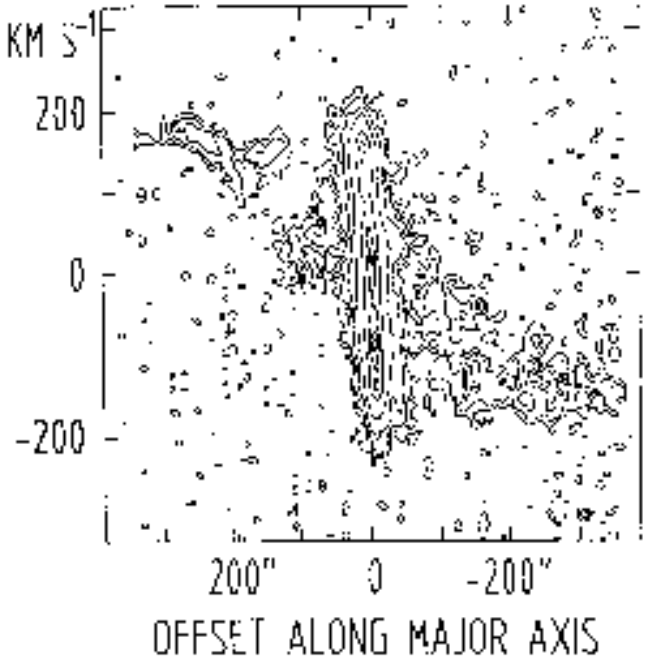,height=9cm}
\caption[CO(2$-$1) pv diagrams] {CO(2$-$1) position-velocity diagram of
NGC\,4945 along the major axis ($PA$ = 43$\degr$) with a resolution of 23$''$
and a channel spacing of 6.3\,\kms. The velocity axis is labeled with respect 
to 560\,\kms. Contours are 75, 150, 300, 600, 1200, 1600, 1800, 1900\,mK on 
a $T_{\rm mb}$ scale. The rms noise is $\sim$50\,mK. For the corresponding 
CO(1$-$0) diagram, see Fig.\,4 of Dahlem et al. (1993).
}
\label{FIG.CO21.PV}
\end{figure}

Fig.\,\ref{FIG.CO21.PV} shows the CO\LTWO\ position-velocity diagram taken
along the major axis. Like Fig.\,\ref{FIG.NUC.HIRES}c, it supports the 
presence of a molecular cloud centred on the nucleus and rotating as a 
solid body with a rotational velocity that reaches $V\sim160$\,\kms. 
The difference between $|V - V_{\rm sys}|$ $\sim$ 160\,\kms\ and the 
edge of the \hbox{\ion{H}{i}} absorption, $|V - V_{\rm sys}|$ $\sim$ 210\,\kms\ 
(Sect.\,3.1.2), is likely reflecting the internal linewidth of molecular 
clouds, the presence of excentric orbits, radial motions caused by 
cloud collisions, or non-axisymmetric distortions of the gravitational
potential (see e.g. Mauersberger et al. 1996b and Sect.\,3.5.5). Radius 
(7\ffas5; taken from Fig.\,\ref{FIG.NUC.HIRES}) and rotational velocity 
($\sim$160\,\kms) are compatible with those obtained for the CO\LONE\ 
and CO\LTRI\ rings observed by Dahlem et al. (1993) and Mauersberger 
et al. (1996a), respectively. Bergman et al. (1992) were the first to 
model the nuclear ring. Henkel et al. (1994) and Curran et al. (2001) 
commented that nuclear gas densities significantly higher than the 
originally proposed value (150\,cm$^{-3}$) are required. Furthermore, 
the central CO\LTWO/CO\LONE\ ratio is a factor of two higher than that 
presumed by Bergman et al. (1992).  

The position-velocity diagrams (Figs.\,\ref{FIG.NUC.HIRES}c and 
\ref{FIG.CO21.PV}) show that for $R$ $<$ 6$''$, the cloud rotates 
like a solid body, but at $R$ $>$ 10$''$ the rotational velocity decreases 
progressively. If this velocity variation is interpreted in terms of 
rotation around a central mass confined essentially within $R$ = 7\ffas5, 
then the estimated mass is 1.5\,10$^9$\,\Msol. This is consistent with a 
value of 0.8--1.6\,10$^9$\,\Msol\ within a radius of $R$ $\sim$ 3$''$, as 
suggested by Koornneef (1993). Mauersberger et al. (1996a) estimated a 
dynamical mass that, scaled to a galaxy distance of 6.7\,Mpc, increases 
from 1.4\,10$^9$\,\Msol\ within $R$ = 5\ffas5 to $\sim$5\,10$^9$\,\Msol\ 
within $R$=33$''$.

Fig.\,\ref{FIG.COMP.PVDIAGRAM}a,b shows \hbox{\ion{H}{i}} position-velocity diagrams 
at two position angles ($49\degr$ and $43\degr$) to best represent the \hbox{\ion{H}{i}} 
motions along the major axis. A noteworthy feature is \hbox{\ion{H}{i}} emission from 
the central region at velocities above $>$200\,\kms\ relative to \Vsys, and 
higher than the velocities seen in \hbox{\ion{H}{i}} absorption or CO emission in 
Fig.\,\ref{FIG.HI.CENABS}. This \hbox{\ion{H}{i}} emission (there is no counterpart
at $V - V_{\rm sys}$ $<$ --200\,\kms) may be related to an optical 
outflow from the centre region towards the halo, first suggested by Nakai 
(1989) on the basis of optical observations. Spectra of the H$\alpha$ and 
[\hbox{\ion{N}{ii}}] line emission close to the nucleus show high velocities consistent 
with a conical outflow of gas (Heckman et al. 1990).

\begin{table}
\caption[Size of nucleus]{Extent of the nuclear CO emission. A gaussian 
beam shape and a 2-dimensional gaussian approximation for the molecular cloud 
distribution have been assumed.
}
\label{TAB.CO.NUCSIZE}
\begin{center}
\begin{tabular}{lclrcl}
\hline
  $^{12}$CO & Apparent & Beam & \multicolumn{2}{c}{Deconvolved} & Ref.\\
  line      & FWHM     &       & \multicolumn{2}{r}{ } &      \\
            &~~$['']$  &$['']$   &~~$['']$  & [pc] & \\
\hline
\LONE   &54$\pm$2 & 43        & 33$\pm$4 & 1100$\pm$130 & 1 \\
\LTWO   &32$\pm$2 & 24        & 21$\pm$4 &  680$\pm$130 & 1 \\
        & 44$\pm$1  & 23       & 37$\pm$4 & 1210$\pm$30  & 2 $^{(a)}$\\
           & 31$\pm$1  & 23       & 21$\pm$4 & 680$\pm$30  & 2 $^{(b)}$\\
\LTRI  &  22$\pm$2        & 15       & 12$\pm$3 & 360$\pm$130  &3 \\
\hline
\end{tabular}
\end{center}
{\footnotesize (1) Dahlem et al. (1993),\\
               (2) this work; $^{(a)}$ major and $^{(b)}$ minor axis values
                   with respect to $PA=13\degr$,\\
               (3) Mauersberger et al. (1996a), adjusted for $D=6.7$\,Mpc
}
\end{table}

\subsection{The overall gas distribution}

Qualitatively, it appears that, in spiral galaxies, neutral atomic gas
dominates in the outer regions, whereas molecular gas is concentrated towards
the inner disk (e.g.\  Sanders et al. 1984). Generally, the CO\LONE\ line
integral is used to estimate the molecular mass content, conventionally
expressed as the \HTWO\ mass. For NGC\,4945, we have used the CO\LTWO\
data on the basis that the uncertainty due to variations in the
CO\LTWO/CO\LONE\ ratio (Fig.\,\ref{FIG.CO21.AT43}b) is considerably less
than the uncertainty in the CO to \HTWO\ conversion factor $X$.

Velocity-integrated \hbox{\ion{H}{i}} and CO line intensities were derived 
from spectra taken along the major axis between $r=100''$ and 360$''$. The 
data indicate that the \hbox{\ion{H}{i}} column density increases, and the 
CO column density decreases, with radius. Using the `standard' conversion 
factor $X$ = 2\x10$^{20}$\,cm$^{-2}$/(K\,\kms) we find for the inner region, 
at e.g.\ $r \sim 135''$, that the column density ratio of \hbox{\ion{H}{i}}/H$_2$ 
is of order 0.2. At the edge of the observed CO\LTWO\ disk, i.e.\ at $R$ $\sim$ 
320 -- 350$''$, the \hbox{\ion{H}{i}}/\HTWO\ column density ratio becomes 
$\sim $1. However, these ratios should be treated with caution, because 
Mauersberger et al. (1996a,b) presented compelling evidence that, at least 
for the central regions of NGC\,253 and NGC\,4945, the value of $X$ is almost 
an order of magnitude lower than the standard value. Thus the 
\hbox{\ion{H}{i}}/H$_2$ abundance ratio should also be $\sim$1 in the inner region.

The velocity-integrated \hbox{\ion{H}{i}} emission (Fig.\,\ref{FIG.MAPS}b) suggests a
variation in major axis position angle and perhaps inclination across the
galaxy. As mentioned earlier, there appears to be an \hbox{\ion{H}{i}} inner disk 
with a lower position angle than that for the outer \hbox{\ion{H}{i}} regions. 
To investigate whether the galaxy is warped, we modeled the \hbox{\ion{H}{i}} 
emission with a set of concentric rings (cf. Begeman 1987; Rogstad et al. 1974) 
of width $50''$ from $R=100''$ to $650''$ and fitted the north-eastern and 
south-western sides of the galaxy separately. The resulting position angles 
and inclinations for the rings both vary over less than 10$^{\circ}$ along 
the major axis. Part of this variation is systematic in that the position angle 
varies from 50$^{\circ}$ in the north-east to 43$^{\circ}$ in the south-west, 
and for $|R|$$>$400$''$ the inclination appears about 3$^{\circ}$ higher in the 
south-west than in the north-east. Significant variations in inclination are 
also found as a function of azimuth and radius in M\,31 (see Braun 1991) and 
may be in part attributable to the presence of spiral arms. 

\subsection{Spiral arms}

It has been pointed out already that the distributions of the 1.4\,GHz continuum, 
\hbox{\ion{H}{i}}, and CO in Fig.\,\ref{FIG.MAPS}, and also the optical image (for
optical images not contaminated by radio contours, see e.g. Nakai 1989; Elmouttie 
et al. 1997) show a common extended emission ridge which is above the major axis 
at positive offsets, and below the major axis at negative offsets. There are many 
emission features associated with the ridge, and the most prominent are listed 
in Table\,\ref{TAB.HI.MAXIMA}. Several emission peaks are located at common major 
axis offsets, and are presumably related physically. The \hbox{\ion{H}{i}} distribution 
shows distinctive features at each end of the galaxy, at offsets of +620$''$ and 
$-580''$. They extend away from the major axis and resemble the tangential locations 
of outer trailing spiral arms of a highly inclined galaxy. At the north-eastern end 
of the galaxy, which is moving away from us, the arm ending at the near side 
extends around behind the plane, above the major axis, and joins an inner disk at the 
location of the emission features at offset $\sim-250''$. At the south-western 
end, rotating towards us, the arm extends around the front of the galaxy, 
below the major axis, joining the inner disk at the location of the emission 
peak at offset +230$''$. The prominent dust lane present in the optical image 
would be associated with this arm. This interpretation accounts for the 
anti-symmetry of the inner emission ridge (see also Dahlem et al. 1993 for 
the inner part of the galaxy).

It is noteworthy that strong IRAS point sources are located at $R\sim240''$ 
on each side of the nucleus. At these locations the 
$^{12}$CO\LTWO/$^{12}$CO\LONE\ line intensity ratio is $>$1. These features 
are consistent with regions of enhanced star formation, which may reside in 
spiral arms viewed near their tangential point.

The interpretation could be extended to include the regions of enhanced 
emission at offsets of about --120$''$ and +120$''$ as additional tangential 
locations of the two spiral arms. Then one arm would extend through offsets 
+620$''$, --250$''$ and +120$''$, and the other through offsets --580$''$, 
+230$''$ and --120$''$.

Interpreting the distributions in Fig.\,\ref{FIG.MAPS} in terms of a 
two-arm spiral structure, we can test whether the designated tangential 
locations are consistent with an overall regular spiral pattern. We 
therefore consider a spiral described by $R=R_0 \times \exp a\vartheta$, 
with the pitch angle $\psi$ in degrees given by $a=\tan \psi$ and the 
azimuth angle $\vartheta$ in radians. Consideration of the two inner 
tangential offsets for each arm yields $\psi=12\degr$ and 13\degr. For 
a pair of positions consisting of the outermost and central offsets, 
$\psi=16$\degr\ for both arms. The results are plausible -- two major 
spiral arms with pitch angles of about $12-13$\degr\ in the inner region 
and $16$\degr\ in the outer region. They are consistent with arms in 
other spiral galaxies (e.g. Puerari \& Dottori 1992) and might also be 
detectable by near infrared imaging.

{\small
\begin{table}[t]
\caption[Emission maxima in the disk]{
Positions of regions with enhanced continuum, \hbox{\ion{H}{i}} or CO emission.
Positional errors are $<$25$''$.}
\label{TAB.HI.MAXIMA}
\begin{center}
\begin{tabular}{ccc}
\hline
\multicolumn{3}{c}{Offsets in arcsec}\\
Cont. & \hbox{\ion{H}{i}} & CO\LTWO
\\
\hline
         & $+$620 &         \\
$+$230   & $+$230 & $+$230   \\
         &        & $+$170   \\
$+$130   & $+$120 & $+$90   \\
$-$55    &        & $-$80    \\
$-$110   & $-$120 & $-$140   \\
$-$175   & $-$185 & $-$195   \\
$-$215   & $-$235 & $-$250   \\
$-$255   & $-$270 &          \\
$-$290   & $-$320 &          \\
         & $-$335 & $-$350   \\
         & $-$580 &          \\
\hline
\end{tabular}
\end{center}
\end{table}
}

\subsection{Disk kinematics and mass}

\subsubsection{Modelling the overall velocity distribution}

A commonly used relation to describe the rotation curve of a galaxy is 
\begin{equation}
\label{EQN.BRANDT}
\frac{V}{V_{\rm max}} =
\frac{\frac{R}{\rvmax}}
{ (\frac{1}{3} + \frac{2}{3}\times(\frac{R}{\rvmax})^n)^{(3/2n)} }
\end{equation}
(see Eqs.\,26 and 28 of Brandt et al. 1960), where $n$ is a measure of the 
steepness of the rotation curve and \Rvmax\ is the radius at which the maximum 
rotation velocity \Vmax\ occurs. This curve results in solid-body rotation at 
small radii and a Keplerian-type velocity decrease at large radii. For a flat 
spiral galaxy inclined to the line-of-sight (such as NGC\,4945), the rotation 
curve can be derived from the variation of radial velocity along the major axis,
assuming that the maximum velocity at any position, corrected for the galaxy 
inclination, represents the tangential (i.e. circular) velocity at the 
corresponding galactocentric radius.

Inspection of the \hbox{\ion{H}{i}} position-velocity diagrams along the major 
axis (Figs.\,\ref{FIG.COMP.PVDIAGRAM}a,b) suggests the existence of two general 
velocity regimes. The rapidly rotating central molecular cloud has been discussed 
already in Sect.\,3.2. The second system relates to the main disk of the galaxy, 
viewed at a position angle of $\sim45^{\circ}$. It extends out to $R\sim400''$, 
with rotational velocities reaching $V\sim170$\,\kms\ (the correction for an 
inclination of $78^{\circ}$ is only $2\%$ and is ignored). This disk is associated 
with the inner concentration seen in the distribution of integrated \hbox{\ion{H}{i}} 
(Fig.\,\ref{FIG.MAPS}b). In the position-velocity diagrams the additional 
\hbox{\ion{H}{i}} concentrations at $R\sim600''$ then correspond to the spiral 
arms that have already been discussed (Sect.\,3.4). 

Because of the high inclination of the galaxy, a line-of-sight at a specific 
offset along the major axis will also include \hbox{\ion{H}{i}} at galactocentric 
radii larger than the radius corresponding to that offset. The line-of-sight 
velocities of the extra \hbox{\ion{H}{i}} components are smaller than the rotational 
velocity at a specific offset and are responsible for the low-velocity `tail' present 
at many positions along the major axis.

Single gaussian profiles were fitted to individual spectra in the \hbox{\ion{H}{i}} 
dataset. To avoid the effects of wings in the \hbox{\ion{H}{i}} spectra, only the 
top half of the \hbox{\ion{H}{i}} profiles with $S_{\rm L}$\,\,$\geq$ $S_{\rm peak}$/2 
was used in the fitting. Applying an iterative procedure, Eq.\,1 was fitted to the 
profile velocities for offsets out to $\pm$450$''$. The parameters listed in Table 
\ref{TAB.HI.BRANDT} were obtained for the disk region. They are consistent with the 
values obtained from a similar fit to CO(1$-$0) data by Dahlem et al. (1993). Our 
$n$ values are 3 (for \hbox{\ion{H}{i}}) and, from a similar fit to the CO(2$-$1) 
data, 4, with small formal errors. Our value for $n$ is significantly smaller than 
that determined by Ables et al. (1987), $n$ $\sim$ 7. This may be a consequence of 
the better angular resolution of our data. Inclusion of the outermost features in 
the analysis would have increased \Rvmax; this value is not well-defined in the fit 
to the \hbox{\ion{H}{i}} position-velocity diagram and the small uncertainty in Table 
\ref{TAB.HI.BRANDT} only accounts for the formal errror. 

{\small
\begin{table}[t]
\caption[Brandt curve model for \hbox{\ion{H}{i}}]{ Parameters derived from the 
Brandt-curve fits and used to model the {\it large scale} velocity field for the 
\hbox{\ion{H}{i}} disk.} 
\label{TAB.HI.BRANDT}
\begin{center}
\begin{tabular}{lll}
& & \hbox{\ion{H}{i}}  \\
\hline
Kinematical centre     & $\alpha (2000)$ & 13$^{\rm h}$ 05$^{\rm m}$ 25\ffs34 \\
                       & $\delta (2000)$ & $-49$\degr$ 29' 09''$  \\
Systemic velocity      & \Vsys  & 557$\pm$3\kms     \\
Max. rotation vel.     & \Vmax  & 170$\pm$5\kms     \\
Turn-over point        & \Rvmax & 380$'' \pm 10$    \\
Position angle         &  $PA$  &  45$\degr\pm 2$   \\
Inclination angle      &  $i$   &  78$\degr\pm 1$   \\
Brandt curve index     &  $n$   &   3.5$\pm$1       \\
\hline
\end{tabular}
\end{center}
\end{table}
}

\subsubsection{The mass of NGC\,4945}

\begin{figure*}
\psfig{figure=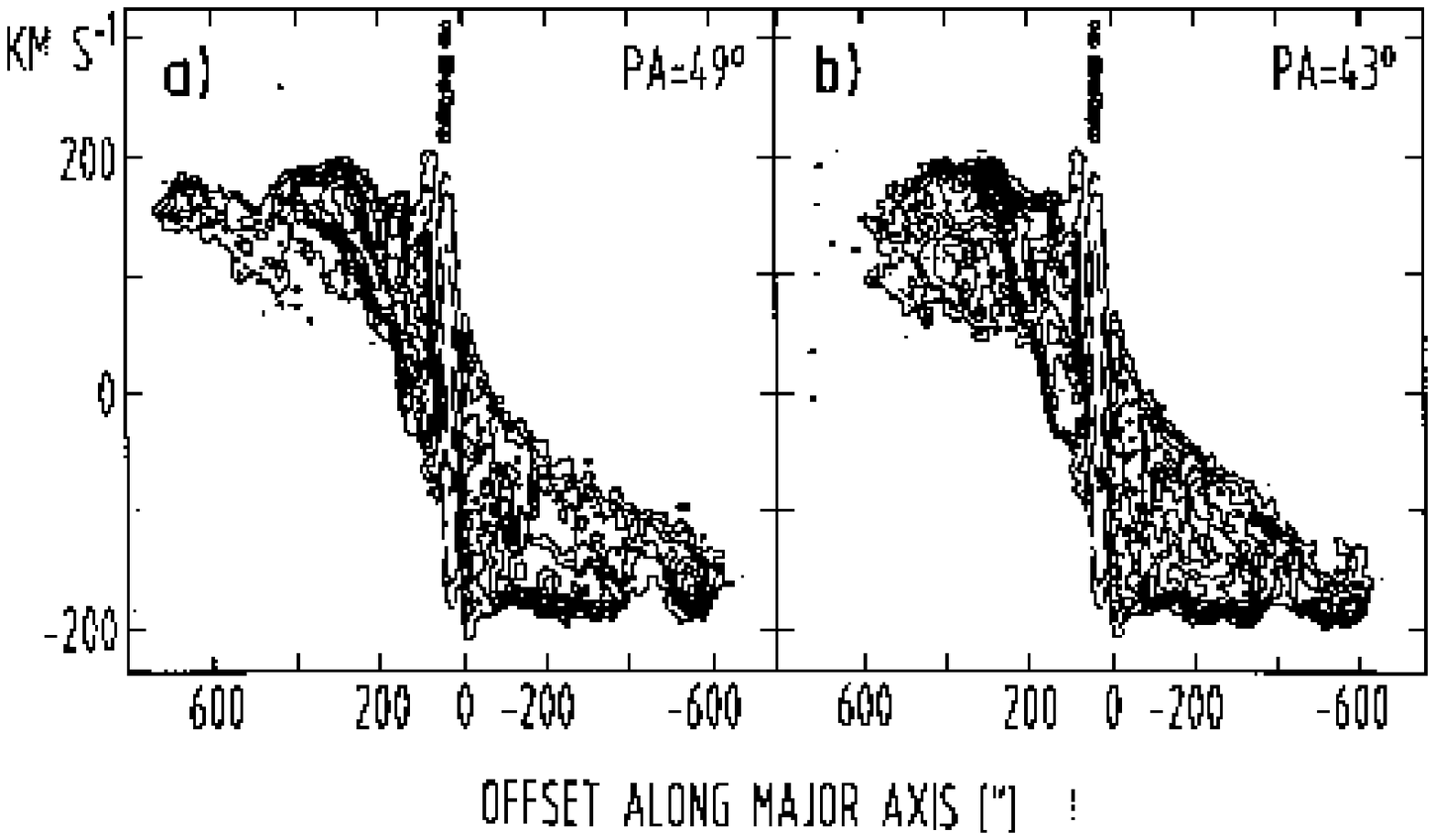,height=9.5cm}
\caption[PV diagrams of \hbox{\ion{H}{i}}]{ Position-velocity diagrams of 
\hbox{\ion{H}{i}} with a resolution of $\sim$23$''$ $\times$ 7.25\,\kms, obtained 
along the major axis at {\bf (a)} $PA=49\degr$ and {\bf (b)} $PA=43\degr$. The 
velocity axis is labeled with respect to 560\,\kms. The contour levels have flux 
densities of -500, -100, -25, 4, 7, 10, 15, 20,  plus {\bf (a)}  30, 40, 50, 55, 
and {\bf (b)}  30, 40, 45\JB.
}
\label{FIG.COMP.PVDIAGRAM}
\end{figure*}

The mass of NGC\,4945 can be derived with 
\begin{equation}
\label{EQN.INT.TOTMASS}
M =(1.5)^{\frac{3}{n}} \times 1242 \times V_{\rm max}^2 \times R_{\rm max}
\times D
\end{equation}
from the parameters in Table \ref{TAB.HI.BRANDT} (see e.g. Eq.\,29 of Brandt
1960; Rogstad et al. 1967; Dahlem et al. 1993). $M$ is in solar masses, \Vmax\ 
is in \kms, \Rmax\ is in arcsec, and $D$ is in Mpc. For $\vmax = 170$\,\kms\ 
and \Rmax = 380$''$, $M$ = 1.4\,10$^{11}$\,\Msol. Reference to the \hbox{\ion{H}{i}} 
position-velocity diagrams (Fig.\,\ref{FIG.COMP.PVDIAGRAM}) suggests that this 
value may be a lower limit, because the velocities of the outermost \hbox{\ion{H}{i}} 
features are larger than the values given by the fitted rotation curve. The paucity of 
\hbox{\ion{H}{i}} at offsets of $400-500''$ may signify a lack of gas at the tangential 
location along the lines-of-sight, in which case the observed velocities are 
less than the rotation velocities (see also Sect.\,3.5.6). 

The total integrated \hbox{\ion{H}{i}} flux density of 70\,Jy\,\kms\ yields an 
\hbox{\ion{H}{i}} mass of 7.5\x10$^{8}$\,M$_{\odot}$, which is only $0.5\%$ of 
the total mass and considerably less than expected for such galaxies. However, 
the presence of the central \hbox{\ion{H}{i}} absorption has badly affected the 
integration. Exclusion of the \hbox{\ion{H}{i}} spectrum observed towards the 
peak of the nuclear continuum emission (see Fig.\,\ref{FIG.HI.CENABS}a) would 
have increased the integrated flux density by more than 200\,Jy\,\kms, yielding 
a total mass fraction of 2\%.  

Because of the limited coverage of the CO distribution, the molecular content 
of the galaxy was not estimated from our CO(2$-$1) data. Assuming an
integrated CO intensity to H$_2$ column density conversion factor of 
$X$ = $N$(H$_2$)/$I_{\rm CO}$ = 2.3\,10$^{20}$\,\cmsq\,(\Kkms)$^{-1}$,
Dahlem et al. (1993) derived a molecular gas mass fraction of 5\% from their 
CO\LONE\ data extending out to 360$''$. While the $X$-factor is appropriate
for the local Galactic disk, it may be too large by a factor of 2--5 for a 
galaxy with a strong nuclear contribution to the overall CO emission (see 
e.g. Mauersberger et al. 1996a). Thus \hbox{\ion{H}{i}} and molecular masses 
in NGC\,4945 should be similar, at the order of 2\% of the total mass inside 
a radius of 380$''$ (12\,kpc).

\subsubsection{Kinematical fine structure}

To compare the \hbox{\ion{H}{i}} and CO kinematics on a small scale, pairs of 
spectra were taken at offsets between  $-100''$ and $-360''$ along the major axis, 
but with a 10$''$ offset to the south; these directions provided the highest CO 
intensities for comparison. Velocity differences between \hbox{\ion{H}{i}} and CO\LTWO\ 
features were found to range from 5 to 20\,\kms. They are not systematic and show 
a random variation in sign. Aside from the existence of \hbox{\ion{H}{i}} and 
CO clouds with separate peculiar motions, for a highly inclined galaxy such 
as NGC\,4945, differences may arise because a number of clouds at different 
physical locations within the galaxy are seen simultaneously in one beam. A 
line-of-sight in a specific direction along the major axis may include clouds 
located at different galactocentric radii or even at different heights above 
and below the galactic plane. 

In the first case (different galactocentric radii), a line-of-sight displacement 
of $1.5$ to $3$\,kpc between \hbox{\ion{H}{i}} and CO could explain velocity 
differences of 5 to 20\,\kms. For comparison, the CO emission peaks in M\,51 
are typically separated by 1.5 to 2\,kpc along the spiral arms (Garc\'{\i}a-Burillo 
et al. 1993). Much smaller line-of-sight displacements could explain the observed 
velocity differences if streaming motions due to density waves were prominent 
(for M\,51, see Aalto et al. 1999).

In the second case (different heights above the plane), significantly 
slower rotation velocities might exist in the halo compared to the 
underlying disk, as observed in the case of M\,82 (Sofue et al. 1992).
Also in our Galaxy peculiar velocities are observed in high-latitude 
\hbox{\ion{H}{i}} gas (e.g. de Boer 1985; their Fig.\,2). 

\subsubsection{Velocity fields and velocity residuals}

\begin{figure*}
\psfig{figure=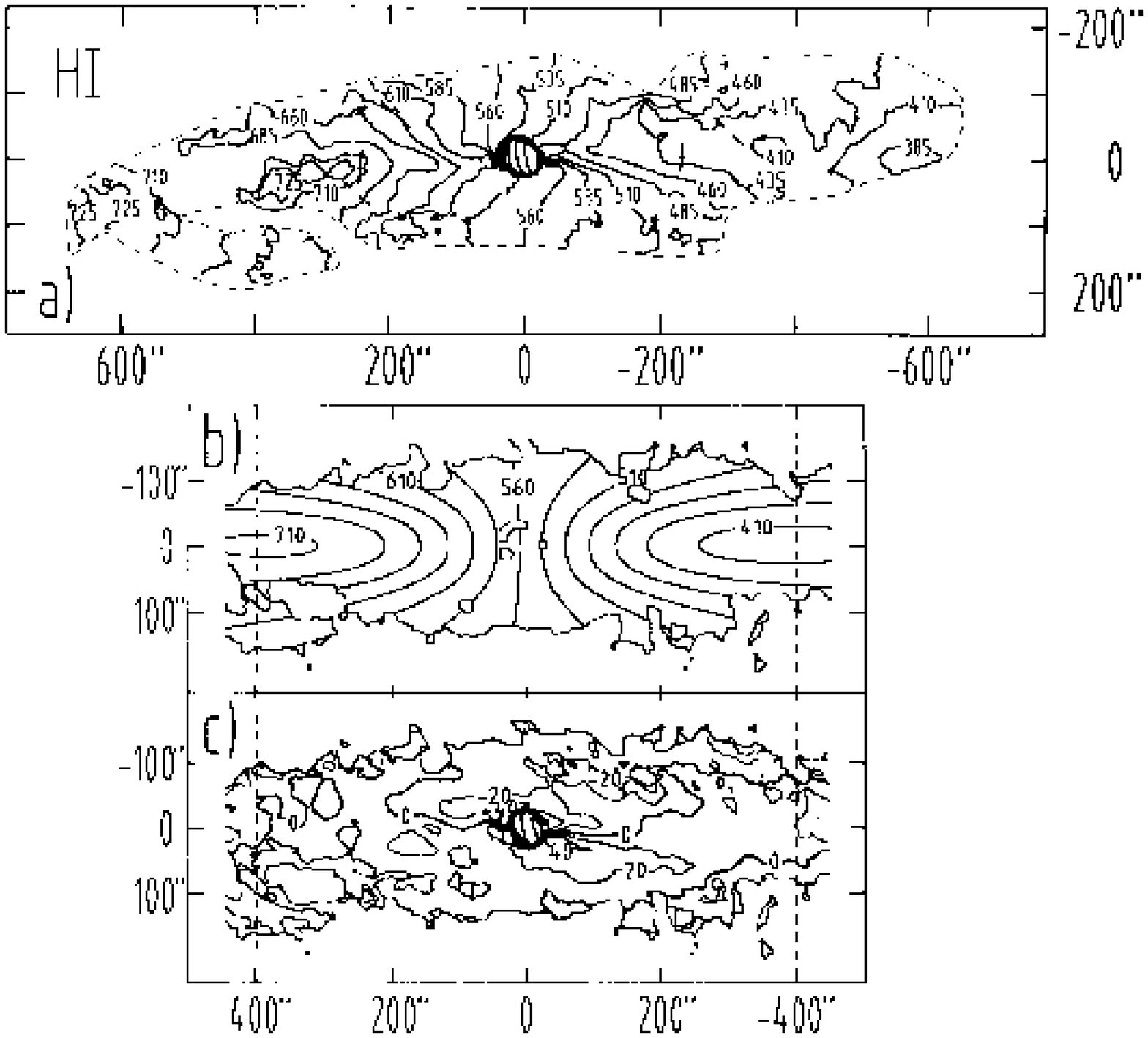,height=17cm}
\caption[Brandt model results.]{ {\bf (a)} \hbox{\ion{H}{i}} velocity field 
obtained from gaussian fitting to observed spectra. The iso-velocity contours 
are at 385, 410, 435, ...\ 710, and 725\,\kms, {\bf (b)} Brandt model velocity 
field for \hbox{\ion{H}{i}}. The contours are spaced at 25\,\kms\ intervals. 
{\bf (c)} Residual velocities, after subtraction of the model field from the 
observed field. The contours are spaced at 20\,\kms\ intervals.
}
\label{FIG.HI.BRANDTMODEL}
\end{figure*}

Fig.\,\ref{FIG.HI.BRANDTMODEL}a shows the \hbox{\ion{H}{i}} velocity field. 
The distribution shows anti-symmetric contour distortions with respect to 
the major axis, indicating departures from uniform circular motion. Considering 
a given velocity isophote north-west of the major axis, contours are displaced towards 
the north, whereas south-east of the major axis contours are displaced to the 
south. The corresponding CO\,(2--1) velocity field indicates a similarly 
anti-symmetric behaviour.

Fig.\,\ref{FIG.HI.BRANDTMODEL}b displays the model velocity field obtained
using Eq.\,1 and the parameters listed in Table \ref{TAB.HI.BRANDT}. To 
examine the departure of \hbox{\ion{H}{i}} motions from the estimated circular 
rotation, the model field was subtracted from the observed \hbox{\ion{H}{i}} field. 
Fig.\,\ref{FIG.HI.BRANDTMODEL}c shows the distribution of \hbox{\ion{H}{i}} residuals. 
In the outer parts of the galaxy the residuals are typically 0--20\,\kms\ with no
systematic trends, indicating that motions are to first order consistent with 
uniform circular motion. However, closer to the nucleus, i.e. above the major 
axis of the galaxy in the north-east and below the major axis in the 
south-west, higher residuals of up to $\pm$40\,\kms\ define an elongated 
region extending across the nucleus to offsets of +120$''$ and $-200''$.
The systematic departures in \hbox{\ion{H}{i}} velocities from circular motion 
were first identified by Ables et al. (1987), but the lower resolution of their 
\hbox{\ion{H}{i}} data (47$''$$\times$73$''$) led to a more widespread contamination 
by the nuclear absorption and inhibited the detection of velocity departures as
large as 40\,\kms. 

Ables et al. (1987) detected velocity residuals of 15 to 30\,\kms\ south-west 
and east and of --15 to --30\,\kms\ west and north-east of the nucleus. This
was interpreted in terms of radial motion towards the nucleus. The residuals
of our fit are slightly different both with respect to amplitude (as already
mentioned) and geometry: In the north-east, north of the major axis, we 
only detect significant negative velocity residuals, while in the south-west, 
south of the major axis, large positive residuals are found  
(Fig.\,\ref{FIG.HI.BRANDTMODEL}c). 

Our Fig.\,\ref{FIG.HI.CENABS} shows a prominent \hbox{\ion{H}{i}} feature displaced 
by about +80\,\kms\ from the systemic velocity. Since this feature is seen in
absorption against the nuclear continuum source, it signifies neutral atomic
gas approaching the galactic centre. From their CO(1$-$0) and (2$-$1) data,
Dahlem et al. (1993) find evidence for inflowing low density molecular gas at
velocities of 80\,\kms. It seems that our nuclear \hbox{\ion{H}{i}} absorption 
spectrum that separates the systemic from the inflowing velocity component 
(for a less conclusive spectrum, see Ables et al. 1987) is tracing the 
neutral atomic part of this kinematical feature as far as it is located in front 
of the nucleus. Note that this gas component is located inside the nuclear 
molecular ring (see Table \ref{TAB.CO.NUCSIZE}; Bergman et al. 1992; Dahlem 
et al. 1993), at a galactocentric radius of a few hundred pc or less. 

\subsubsection{Evidence for a bar}

De Vaucouleurs (1964) suggested that NGC\,4945 contains a central bar, which
provides a very efficient mechanism to quickly transport matter towards the
nucleus while releasing angular momentum outwards. Can such a bar explain the 
velocity anomalies outlined in Sect.\,3.5.4? A classical signature of a bar is 
an S-shaped distortion in the isovelocity contours of the gas (e.g. Kalnajs 
1978; Duval \& Monnet 1985). This is observed in NGC4945 (see 
Fig.\,\ref{FIG.HI.BRANDTMODEL}a; Sect.\,3.5.4). If a bar is present, the 
\hbox{\ion{H}{i}} velocity residuals in Fig.\,\ref{FIG.HI.BRANDTMODEL}c could 
be interpreted in terms of systematic overall gas flows along it, approaching 
at positive offsets, receding at negative offsets. The orbit velocities in the 
bar near the nucleus may be estimated from the pv diagrams and the velocity field 
residuals. The continuum and CO distributions in Figs.\,\ref{FIG.MAPS} 
and \ref{FIG.CO21.AT43} show enhanced emission extending out from the 
nucleus along $PA$ $\sim$ 33$^{\circ}$, i.e. along the axis that is also
showing the velocity anomalies (Fig.\,\ref{FIG.HI.BRANDTMODEL}c). Assuming
that the putative bar is associated with the emission ridge introduced
above and that this is located within the plane of the galaxy, we obtain
an azimuthal angle of $\sim40\degr$ (counterclockwise) with respect to 
the line-of-sight projected onto the plane of the galaxy. The total inclination
to the line-of-sight would then be 45$^{\circ}$. 

There exists a strong correlation between the nuclear absorbing column density
and the presence of a bar in Seyfert 2 galaxies. Strongly barred Seyfert 2 
galaxies have an average $N_{\rm H}$ that is two orders of magnitude higher 
than in non-barred Sy 2s. More than 80\% of the `Compton thick Seyfert 2s'
($N_{\rm H}$ $\ga$ 10$^{24}$\,\cmsq; most of this column density must arise 
from the innermost few 10\,pc) are barred and almost 60\% of these are 
`strongly' barred (Maiolino et al. 1999). NGC\,4945 is a Compton thick Seyfert 
2 galaxy (e.g. Guainazzi et al. 2000; Madejski et al. 2000).

Non-interacting spiral galaxies with moderate inclination and substantial far 
infrared emission (for the central part of NGC\,4945, \irlum $\sim$ 
5\,10$^{10}$\,\solum; IRAS 1989) are known to show both {\it strong and long 
bars} (Martinet \& Friedli 1997). The presence of asymmetric morphologies
in individual Seyfert galaxies is positively correlated with their tendency
to exhibit enhanced star forming activity (Maiolino et al. 1997). Hence 
irrespective of direct observational evidence, the presence of a bar in 
NGC\,4945 is, at least statistically, very likely. 

The velocity dispersion of \hbox{\ion{H}{i}} in the central region is also 
consistent with the presence of a bar. Estimated for each observed 
\hbox{\ion{H}{i}} spectrum during the gaussian fitting, the disperson 
in the outer galaxy is 10 -- 30\,\kms, which is consistent with the 
dispersion of gas along the spiral arms in the presence of relatively 
uniform circular rotation. However, within $\sim150''$ of the nucleus 
there is an elongated region where the dispersion is above 50\,\kms. 
This may reflect higher gas turbulence or a higher space density of gas 
clouds near the centre, but could also reflect fast motion on highly 
eccentric orbits in a bar. Since all this evidence is circumstantial, 
a definite answer to the question whether NGC\,4945 has a bar must come 
from the near-infrared (preferably K-band) image. This method proved 
to be conclusive in the case of another nearby highly inclined southern 
starburst galaxy, NGC\,253 (see e.g. Engelbracht et al. 1998).  

\subsubsection{Dynamical analysis}

The \hbox{\ion{H}{i}} and CO\LTWO\ pv diagrams can be analysed using 
linear resonance theory (e.g. Binney \& Tremaine 1987) to deduce the 
locations of various gravitational resonances within NGC\,4945. If we 
assume the presence of a `weak' bar (the orbits can be described by a 
superposition of circular motion around the centre and small oscillations 
caused by the non-axisymmetric part of the potential), the spiral and 
barred structure is constrained by the locations of these resonances. 
For a perfectly edge-on barred spiral the locations can be only approximate, 
because the rotation curve will be affected by the resonances as well as 
the bar's de-projected size and angle to the line-of-sight. In the case of 
NGC\,4945, the proposed bar (Sect.\,3.5.5) is sufficiently displaced from 
the major axis to provide a reasonable approximation. 

\begin{figure}
\psfig{figure=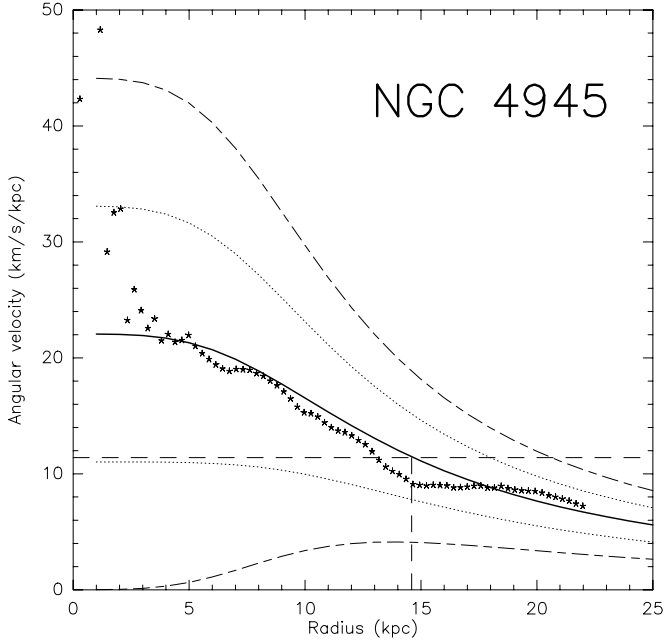,height=9.0cm,angle=-90}
\caption[velocity/radius diagram]{Variation of angular velocity ($\Omega$)
as a function of galactocentric radius, determined from \hbox{\ion{H}{i}} observations 
(asterisks; see Sect.\,3.5.1 for details) and fitted with a Brandt rotation 
curve (solid line; see Eq.\,1) for the outer disk with parameters given in 
Table \ref{TAB.HI.BRANDT}. At $R$ $\sim$ 15\,kpc, low measured angular 
velocities with respect to the fit are likely an artefact caused by a lack 
of neutral atomic gas at this radius. Shown are also $\Omega-\kappa/m$ and 
$\Omega+\kappa/m$ for $m=2$ (dashed lines) and $m=4$ (dotted lines). The 
assumed co-rotation radius (vertical dashed line) and resulting pattern 
speed $\Omega_{\rm P}$ (horizontal dashed line) are indicated.
}
\label{FIG.VRAD.DYN}
\end{figure}

For gas orbiting the centre of a galaxy with angular velocity $\Omega = 
V/R$ at radius R, the radial epicyclic frequency $\kappa$ is expressed by
\begin{equation}
\label{EQN.DYN.FIT}
\kappa (R) = \sqrt{4\,\Omega^2(R) + R\,\frac{{\rm d}\Omega^2(R)}{{\rm d}R}}.
\end {equation}
Application of a Brandt rotation curve (Eq.\,1) infers
\begin{eqnarray}
\frac{{\rm d}\Omega^2}{{\rm d}R} = -\frac{2\,V_{\rm max}^2}{R_{\rm V_{max}}^3}
\left (\frac{R}{R_{\rm V_{max}}}\right )^{n-1}
\left [\frac{1}{3} + \frac{2}{3}\left 
(\frac{R}{R_{\rm V_{max}}}\right )^n\right ]^{\frac{-3-n}{n}} . \nonumber \\
\end{eqnarray}
Inner and outer `Lindblad' resonances (ILR and OLR) occur when the pattern
speed of the bar $\Omega_{\rm P} = \Omega-\kappa/m$ and $\Omega+\kappa/m$, 
respectively ($m$=2 represents a barred potential, see e.g. Eq.\,3--115 of Binney 
\& Tremaine 1987). Fig.\,\ref{FIG.VRAD.DYN} shows the measured variation of angular 
velocity as a function of galactocentric radius. A rotation curve of the 
form given in Eq.\,1 with the parameters outlined in Table \ref{TAB.HI.BRANDT} 
has been fitted to \hbox{\ion{H}{i}} velocities derived from the \hbox{\ion{H}{i}} 
pv diagram in Fig.\,\ref{FIG.COMP.PVDIAGRAM}. The fitted curve provides a 
reasonable fit to the outer galaxy ($R$ $\ga$ 5\,kpc).

For a barred spiral galaxy, it is expected that the co-rotation radius CR
would be located not far beyond the end of the bar. For NGC\,4945 this 
position is uncertain. Accounting for unsystematic velocity residuals 
of 0 -- 20\,\kms\ (Sect.\,3.5.4), the velocity anomaly can be traced out 
to approximately $\pm$150$''$ projected on the sky (see 
Fig.\,\ref{FIG.HI.BRANDTMODEL}c). With an inclination to the line-of-sight 
of 45$^{\circ}$ (Sect.\,3.5.5) this would be equivalent to a bar extent 
(from the nucleus) of about 7\,kpc. Since the spiral arms could be 
traced from the outer galaxy ($\sim$600$''$ offset from the centre) 
to $\sim$120$''$ (Sect.\,3.4), the spiral arms might 
be connected with the bar, as commonly observed in galaxies with small
inclination (e.g. Reynaud \& Downes 1997; H{\"u}ttemeister et al. 1999). 
The CR must be located at $R$$>$7\,kpc, the outer radius of the gaseous bar. 
Since stellar bars are longer than their gaseous counterparts (Martinet 1995) 
and since gravitational torques de-populate the co-rotation region in a spiral 
galaxy (e.g. Garc\'{\i}a-Burillo \& Gu{\'e}lin 1995; Combes 1996), the drop in 
\hbox{\ion{H}{i}} intensity seen in Fig.\,\ref{FIG.COMP.PVDIAGRAM} at $R$ $\sim$ 450$''$ 
(14.6\,kpc) and the small `rotational' velocities at this radius (see 
Fig.\,\ref{FIG.VRAD.DYN} and Sect.\,3.5.2) may be considered as a signature 
of the CR (see also Freeman 1997 for `average' radii). Adopting this radius, 
the pattern speed is 11\,\kms\,kpc$^{-1}$; the OLR (which occurs where the 
pattern speed intercepts the upper $m$=2 curve in Fig.\,\ref{FIG.VRAD.DYN}) 
is then at a radius of 21\,kpc (angular distance: $\sim$650$''$) and the 
$m$=4 ultra-harmonic resonance is at 18\,kpc (550$''$), where the outermost 
\hbox{\ion{H}{i}} features are seen (Fig.\,\ref{FIG.MAPS}).


Our Brandt rotation curve is a good fit to the outer galaxy, but does not
allow to predict the presence of an ILR. In the innermost parts of the
galaxy, measured angular velocities are larger than those suggested by the 
Brandt curve (Fig.\,\ref{FIG.VRAD.DYN}), leaving open the possibility that
an inner Lindblad resonance exists at $R$ $<$ 3\,kpc. An attractive 
but speculative view is to associate the inner molecular ring at a 
galactocentric radius of a few hundred pc (Table \ref{TAB.CO.NUCSIZE}) 
with the ILR that might contain a nested secondary bar (the 80\,\kms\ inflow; 
see Sect.\,3.5.4) guiding atomic and molecular gas to the putative 
circumnuclear torus discovered by Greenhill et al. (1997). Note however 
that the nuclear molecular ring is part of the transitional region between 
solid body and differential rotation. The ring is therefore not necessarily 
formed by a gravitational resonance. It could also be caused by viscous 
transport as e.g. outlined by D{\"a}ther \& Biermann (1990). Adopting the 
D{\"a}ther \& Biermann mechanism, an age estimate applying their last equation 
leads to a formation timescale of $t_{\rm ring}$ $\sim$ 1\,Gyr.

\section{Conclusions}

A high-resolution study has been performed of the 1.4\,GHz continuum, 
\hbox{\ion{H}{i}}, and CO\LTWO\ emission for the southern spiral galaxy 
NGC\,4945. It utilizes both the Australia Telescope Compact Array (ATCA) 
and the Swedish-ESO-Submillimetre Telescope (SEST). The angular resolution is
$\sim$23$''$ (750 pc at $D$=6.7\,Mpc) and the spectral resolution is
$\sim$7\,\kms. The ATCA results also yield high resolution ($\sim$3.6$''$) 
images of the nuclear region of the galaxy.

The main conclusions are as follows:
\begin{enumerate}

\item
The 1.4\,GHz continuum emission extends over 11.6$'$$\times$3.3$'$ with 
$PA\sim45\degr$ and contains a bright nucleus with a size of 
7\ffas 6$\times$3\ffas 4$\,(\pm0\ffas 2)$ (corresponding to 
250$\times$110\,pc) with the major axis at $PA=42\fdg 5$ and an integrated flux 
density of $4.6\pm0.1$\,Jy, centred at $\alpha(2000) = 13^{\rm h} 05^{\rm m}$ 
27\ffs3, $\delta(2000) = -49\degr 28' 07''$. 

\item
Our CO\LTWO\ data confirm the presence of a circumnuclear ring at $R$ $\sim$
10--20$''$. \hbox{\ion{H}{i}} gas is observed near the centre with a rotation 
velocity that reaches 160\,\kms\ at $R$=7\ffas5 (250\,pc); at smaller and larger 
radii the velocity is lower. The dynamical mass inside of 7\ffas5 is 
1.5\,10$^9$\,\Msol.

\item
Nuclear \hbox{\ion{H}{i}} emission at velocities offset by more than 200\,\kms\ 
from the systemic velocity may be related to an outflow towards the halo, observed 
at optical wavelengths. \hbox{\ion{H}{i}} absorption shows evidence for inflowing 
nuclear gas ($|V - V_{\rm sys}|$ $\sim$ +80\,\kms) that was previously only seen
in CO.

\item 
\ion{H}{i} is detected beyond the optical 25$^{\rm m}\,$arcsec$^{-2}$ contours. 
The \hbox{\ion{H}{i}} emission extends over $\sim22'$ (43\,kpc) at a position angle 
that increases from $43\degr$ to $50\degr$ from the south-west to the north-east. 
\hbox{\ion{H}{i}} Local Standard of Rest velocities extend from 340 to 770\,\kms\ 
and yield a systemic velocity of 555--560\,\kms. An analysis of the velocity field 
indicates that the maximum mean rotation velocity of 170\,\kms\ occurs at a 
galactocentric radius of $\ga$380$''$, yielding a total mass of 
1.4\,$10^{11}$\,\Msol\ inside of $R$ = 380$''$. Mass fractions of molecular 
and \hbox{\ion{H}{i}} gas are $\sim$2\% for each component.

\item
The \hbox{\ion{H}{i}} distribution shows distinctive features at each end of 
the major axis, offset about $\pm600''$ from the nucleus. These are interpreted 
as spiral arms viewed tangentially. Other common features in the continuum, 
\hbox{\ion{H}{i}} and CO distributions at offsets near $\pm120''$ and $\pm240''$ 
may be additional tangential directions for the arms. The results are consistent 
with two spiral arms having pitch angles of 12-13\degr\ in the inner ($|R|$$<$240$''$) 
and 16\degr\ in the outer galaxy. 

\item
The \hbox{\ion{H}{i}} velocity field of the galaxy shows a pattern generally 
consistent with uniform circular motion. Near the nucleus, however, systematic 
departures exceeding $\pm30$\,\kms, S-shaped velocity contours, high infrared 
luminosities, large nuclear gas column densities, and increased velocity dispersions 
are interpreted in terms of a bar extending across the nucleus. The data 
suggest a structure with $PA$ $\sim$ 35$^{\circ}$, an azimuthal angle of 
$\sim$40$^{\circ}$ (counterclockwise), extending to radial offsets of 
$\pm150''$ and connecting the inner spiral arms. The likely presence of a bar 
and large amounts of molecular gas in the inner disk strongly suggest 
that the starburst in NGC\,4945 is ongoing and that a post-starburst 
stage of evolution is not yet reached. 

\end{enumerate}

\begin{acknowledgements}
This project was partly supported by the Max-Planck Forschungspreis 1992,
awarded to JBW and RW. Discussions with S. H{\"u}ttemeister and detailed
comments by an anonymous referee are gratefully acknowledged.
\end{acknowledgements}

{}

\end{document}